\documentclass[aps,pre]{revtex4-1}

\usepackage{amsmath, amsthm, amssymb}
\usepackage{graphicx}

\begin{document}


\title{On the entropy of plasmas described with regularized $\kappa$-distributions}


\author{H.~Fichtner, K.~Scherer, M.~Lazar}
\affiliation{
Institut f\"ur Theoretische Physik IV, 
Ruhr-Universit\"at Bochum,
44780 Bochum, 
Germany
}
\author{H.J.~Fahr} 
\affiliation{
Argelander Institut f\"ur Astronomie,
Universit\"at Bonn,
Auf dem H\"ugel 71,
53121 Bonn,
Germany
}
\author{Z.~V\"or\"os}
\affiliation{
Space Research Institute, 
Austrian Academy of Sciences,
Schmiedlstr. 6, 
8042 Graz,
Austria
}

\date{\today}

\begin{abstract}
In classical thermodynamics the entropy is an extensive quantity, i.e.\ 
the sum of the entropies of two subsystems in equilibrium with each other 
is equal to the entropy of the full system consisting of the two subsystems.
The extensitivity of entropy has been questioned in the context of a 
theoretical foundation for the so-called $\kappa$-distributions, which 
describe plasma constituents with power-law velocity distributions. 
We demonstrate here, by employing the recently introduced {\it regularized 
$\kappa$-distributions}, that entropy can be defined as an extensive 
quantity even for such power-law-like distributions that truncate exponentially.  
\end{abstract}


\maketitle


\section{Introduction and Motivation}
The so-called $\kappa$-distributions have become popular \cite[e.g.,][]{Pierrard-Lazar-2010} to quantitatively describe the power law 
behaviour of velocity, momentum or energy distributions of various energetic particle populations, 
reaching from flare-accelerated electrons \cite[e.g.,][]{Kasparova-Karlicky-2009, Oka-etal-2013, Effenberger-etal-2017}
via suprathermal electrons and ions in the interplanetary medium 
\cite[e.g.,][]{Fisk-Gloeckler-2012, Maksimovic-etal-2005, Stverak-etal-2009, Vocks-2012} as 
well as in the outer heliosphere \cite{Fahr-etal-2016, Fahr-etal-2017} even to laboratory laser physics
\cite[e.g.,][]{Webb-etal-2012, Elkamash-Kourakis-2016}. These distributions 
have been employed in most cases as useful tools, i.e., in the pragmatic spirit with which they were 
introduced 50~years ago in the context of magnetospheric physics \cite{Olbert-1968, Vasyliunas-1968}. 

Attempts to physically justify these special power laws can be divided into two groups. On the one 
hand, it is possible to rigorously derive $\kappa$-distributions for specific systems where 
particles interact with external radiation \cite{Hasegawa-etal-1985}, with plasma fluctuations
\cite[e.g,][]{Ma-Summers-1998, Shizgal-2007, Yoon-2014, Effenberger-etal-2017}, or with a constant 
temperature heat bath \cite{Shizgal-2018}. On the other hand, 
$\kappa$-distributions should be motivated on the basis of more fundamental considerations related 
to generalizations of the concept of entropy \cite{Tsallis-1988} or Gibbsian theory
\cite{Treumann-Jaroschek-2008}. Both approaches face limitations: While the former appears to be
valid for systems with special constraints resulting in a special class of $\kappa$-distributions 
(termed `Kappa~A', see below) and only allows specific $\kappa$-values as discussed in 
\cite{Lazar-etal-2016}, the latter is requiring a generalized, non-extensive entropy, apparently 
implying internal inconsistencies \cite{Nauenberg-2003} that have as yet not been resolved 
\cite{Tsallis-2004, Nauenberg-2004}. 

It has been pointed out recently by Scherer et al.\ \cite{Scherer-etal-2017} that, even if these difficulties
could eventually be overcome, the resulting $\kappa$-distributions would still be hampered by an 
only finite number of non-diverging velocity moments, i.e.\ the condition that $\kappa>(l+1)/2$ for the
velocity moment of order $l$ to exist. This implies, in particular, that the definition of the 
$\kappa$-distribution itself, requiring the existence of the second-order moment, i.e., kinetic temperature, 
is valid only for $\kappa> 3/2$. Moreover, the heat flux is given by the third-order moment and requires
even larger values $\kappa> 5/2$ \citep[see, e.g.,][]{Shaaban-etal-2018a,Shaaban-etal-2018b},
while the convergence of higher-order moments should ensure closure schemes for a macroscopic description.
These motivated  Scherer et al.\ \cite{Scherer-etal-2017} to introduce the {\it regularized 
$\kappa$-distribution} (RKD). The suggested regularization removes all divergences, allows to analytically 
calculate all (isotropic) velocity moments for all positive $\kappa$-values, and may adjust to power-law distributions 
observed in the solar wind with clear evidences of exponential cutoffs.
As we demonstrate in the present paper, these improvements are not the only
advantages: the RKD also possesses an additive entropy, which is, thus, an extensive quantity.  

The paper is organized as follows. In the next two sections~\ref{sec-kappa-dist} and~\ref{bg-entropy} we define 
the different forms of the $\kappa$-distributions discussed in the literature, consider the (Boltzmann-)Gibbs
entropy, and calculate the entropy of a spatially homogeneous RKD plasma.
In section~\ref{isolated} we demonstrate explicitly, with an application to isolated plasmas,
the additivity, i.e., extensitivity of the RKD's entropy. 
Finally, in section~\ref{sec-ihs}, we consider an inhomogeneous system that is better described with the entropy
density rather than the entropy itself. All results are summarized and discussed in a concluding 
section~\ref{conclusions}. 

\section{$\kappa$-distributions: definitions}
\label{sec-kappa-dist}
Most of the applications and fundamental approaches considering (isotropic) $\kappa$-distributions
employ the following form, originally introduced in \cite{Olbert-1968, Vasyliunas-1968},
\begin{eqnarray}
f_\kappa (v) = & \displaystyle \frac{n}{\pi^{3/2}\Theta^3} \, 
                 \frac{\Gamma[\kappa +1]}{\kappa^{3/2}\Gamma[\kappa-1/2]} 
                 \left(1 + \frac{v^2}{\kappa\Theta^2}\right)^{-\kappa-1}, 
\label{kappa-dist}
\end{eqnarray}
where $n$ denotes the number density of the considered particle species, $\Gamma[x]$ the gamma function, 
$v$ the particle speed, and $\kappa > 3/2$. The reference speed $\Theta$, introduced as the most probable speed 
\cite{Vasyliunas-1968}, is related to a kinetic temperature $T$ via the second-order moment
\begin{eqnarray}
T = \frac{m}{n k_B} \int v^2 f_\kappa(v) d^3 v 
  = \frac{\kappa}{\kappa -3/2} \, \frac{m}{2 k_B} \Theta^2 
\label{kappa-temp}
\end{eqnarray}
Here, $m$ is the particle mass and $k_B$ the Boltzmann constant. 
For a generalization to bi-$\kappa$-distributions, see, e.g., \cite{Summers-Thorne-1991} and 
\cite{Lazar-etal-2015}. 

\begin{figure}
\includegraphics[width = 0.75\columnwidth]{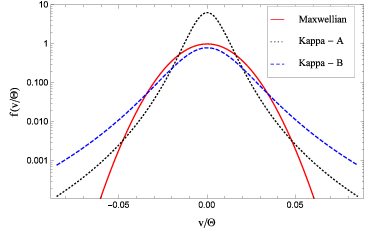}
\caption{The Kappa-A and Kappa-B distributions discussed in section~II in comparison to the Maxwellian obtained 
         in the limit $\kappa\to\infty$. Adapted from \cite{Lazar-etal-2015}.} 
\label{kappas}
\end{figure}

One distinguishes two choices. The first is to consider the temperature in Eq.(\ref{kappa-temp}) 
to be always equal to that of the associated Maxwellian \cite[e.g.,][]{Hellberg-etal-2009, Yoon-2014, Astfalk-etal-2015,
Livadiotis-2017},
\begin{eqnarray}
f_M(v) = \frac{n}{(\sqrt{\pi} v_{th})^3} \exp\left(-\frac{v^2}{v_{th}^2}\right),
\label{max-dist}
\end{eqnarray}
which enables to extend the concept of temperature of a $\kappa$-distribution in the strict sense of thermodynamics. 
This choice naturally implies a distribution that is not only above the associated Maxwellian at high but also
at low speeds, see `Kappa~A' in Figure~\ref{kappas}.
The alternative is obtained in the limit $\kappa\to\infty$, when the speed $\Theta$ is independent of $\kappa$ 
(and  equals the thermal speed $v_{th}$ of Maxwellian limit) allowing for the modelling of
suprathermal wings of a distribution, denoted `Kappa~B' in Figure~\ref{kappas}, on the expense of its
core population \cite[e.g.,][]{Olbert-1968, Hau-Fu-2007, Fahr-etal-2014, Lazar-etal-2016}, i.e.\ without any 
enhancement at low speeds relative to the associated Maxwellian. 
While systems properly described with Kappa~A are usually specifically set up and consistent with special
(isolated) $\kappa$-values, and exhibit $\kappa$-dependent speeds $\Theta=\Theta_\kappa$, those obeying Kappa~B 
have less constraints \cite{Hau-etal-2009} and describe total populations with temperatures increasing for 
decreasing $\kappa$-value. For the ongoing debate about which choice is correct or, at least, represents an
appropriate description of a given system, see \cite{Lazar-etal-2016} and \cite{Livadiotis-2017}.

It must be noted that both choices exhibit unphysical features. First, in the usual classical (as opposed to 
a relativistic) treatment the power law (\ref{kappa-dist}) extends to infinite speeds, implying an infinite
number of diverging velocity moments. While this feature has been tried to be explained some while ago in the
context of a finite sample size effect and the concept of self-organized criticality \cite[e.g.][]{Schertzer-Lovejoy-1993},
regarding the $\kappa$-distributions there is a second unphysical feature, namely that even formally existing moments are
diverging for values 
$\kappa \leq 3/2$, see, for example, Eq.(\ref{kappa-temp}). In order to remove these unphysical features
Scherer et al.\ \cite{Scherer-etal-2017} defined the {\it regularized $\kappa$-distribution} (RKD)
\begin{eqnarray} 
f_{RKD}(v) = n\, A \left(1 + \frac{v^2}{\kappa\Theta^2}\right)^{-\kappa-1}
             \exp\left(-\alpha^2\frac{v^2}{\Theta^2}\right) \equiv n\, g_{RKD}(v),
\label{rkd-dist}
\end{eqnarray} 
by introducing a physically motivated exponential cut-off controlled via the parameter $\alpha$. 
For sufficient low values of the latter both the low-order velocity moments and the kinetic 
properties are virtually the same as for the corresponding standard $\kappa$-distributions. 
While, again in view of a finite sample size effect, it might be difficult to determine the 
value of this cut-off parameter in all cases, examples for such determination can be found in 
\cite{Scherer-etal-2017}. Most importantly, the RKD allows an analytical 
calculation of all velocity moments for all positive $\kappa$-values.  
$A=A(\kappa,\alpha,\Theta)$ is the required normalization constant.
\section{Gibbs Entropy}
\label{bg-entropy}
A general definition of entropy $S$ that is valid both for equilibrium and non-equilibrium systems
was given originally by Boltzmann \cite{Boltzmann-1872} and Gibbs \cite{Gibbs-1902} and specifically for a plasma
constituent more recently, e.g., by Balescu \cite{Balescu-1975, Balescu-1988}, and
Cercignani \cite{Cercignani-1988}:
%
\begin{eqnarray} 
S = - k_B \iint f\, [\ln(f)-1]\; d^3\!r d^3\!v - k_B N \ln\left(\frac{h^3}{m^3}\right)
\label{entropy}
\end{eqnarray} 
where $f=f(\vec{r},\vec{v},t)$ is the phase space distribution function of $N$ particles of the considered 
species and $h$ is the Planck constant.
This definition of the Gibbs entropy (sometimes called Boltzmann-Gibbs entropy) not only takes into account the quantum mechanical lower limit of the phase space volume
occupied by a single particle and also contains the Gibbs factor in order to avoid the Gibbs paradoxon
\cite{Gibbs-1902} that is related to the indistinguishability of states after interchanging identical
particles, it is also valid for non-equilibrium systems \cite{Balescu-1988, Cercignani-1988}. While this
definition is useful for the case of homogeneous plasmas, it is more appropriate to define an entropy
density for spatially inhomogeneous systems, an example of which we discuss in section~\ref{sec-ihs}.

In the following, we first briefly review the calculation of the entropy for a Maxwellian plasma 
constituent and then apply the above definition for a plasma constituent obeying an RKD 
\cite{Scherer-etal-2017}. For both cases we assume stationary, isolated plasmas with vanishing spatial
gradients, i.e., we assume spatial homogeneity.
\subsection{Maxwellian plasma}
The (non-drifting) Maxwellian distribution is given by Eq.(\ref{max-dist}). With the above assumptions neither
$n$ nor $T$ are a function of the location $\vec{r}$. Using this distribution in Eq.(\ref{entropy}) leads to 
\begin{eqnarray} 
S_M &=& -k_B \iint f_M\, [\ln(f_M)-1]\; d^3\!r d^3\!v 
        -k_B N \ln\left(\frac{h^3}{m^3}\right)\nonumber\\
    &=& -k_B \, \ln\left(\frac{n}{(\sqrt{\pi} v_{th})^3}\right) \iint f_M\; d^3\!r d^3\!v
        +\frac{k_B}{v_{th}^2} \iint f_M v^2\; d^3\!r d^3\!v\nonumber\\ 
    & & +k_B \iint f_M\; d^3\!r d^3\!v 
        -k_B N \ln\left(\frac{h^3}{m^3}\right)
\end{eqnarray} 
With the usual definition of the zeroth- and second-order moments
\begin{eqnarray} 
&& N = \int n \; d^3\!r = \iint f_M\; d^3\!v d^3\!r \\
&& T = \frac{m}{3 k_B n} \int f_M v^2 d^3\!v 
\end{eqnarray} 
the Maxwellian entropy can be written as 
\begin{eqnarray} 
S_M = -k_B N \, \ln\left(\frac{n h^3}{(2 \pi m k_B T)^{3/2}} \right) + \frac{5}{2}\,k_B N 
\end{eqnarray} 
which, upon introducing the so-called thermal de~Broglie wavelength $\lambda = h/\sqrt{2 \pi m k_B T}$
\cite[e.g.,][]{Goeke-2010}, reads 
\begin{eqnarray} 
S_M = k_B N\,\left[\ln\left(\frac{1}{n \lambda^3}\right) + \frac{5}{2}\right] 
\end{eqnarray} 
As long as the above assumption of constant number density $n$ holds, $S_M$ is proportional to 
the total number of particles $N$ and, thus, it is an extensive quantity. 
\subsection{RKD plasma}
The (non-drifting) RKD \cite{Scherer-etal-2017} is given 
in Eq.(\ref{rkd-dist}). As before, all quantities are assumed to be independent of location. 
Note that the phase space distribution $f_{RKD}$ is normalized to $n$ while the velocity 
distribution $g_{RKD}$ is normalized to unity. Using the RKD in Eq.(\ref{entropy}) leads to 
\begin{eqnarray} 
S_{RKD} &=& -k_B \iint f_{RKD}\, [\ln(f_{RKD})-1]\; d^3\!r d^3\!v 
            -k_B N \ln\left(\frac{h^3}{m^3}\right)\nonumber\\
        &=& -k_B \, \ln(n A) \iint f_{RKD}\; d^3\!r d^3\!v\nonumber\\
        & & -k_B \iint f_{RKD}\, \ln\left(1 + \frac{v^2}{\kappa\Theta^2}\right)^{-\kappa-1} d^3\!r d^3\!v
         \nonumber\\
        & & +k_B \frac{\alpha^2}{\Theta^2} \iint f_{RKD} v^2\; d^3\!r d^3\!v\nonumber\\ 
        & & +k_B \iint f_{RKD}\; d^3\!r d^3\!v 
            -k_B N \ln\left(\frac{h^3}{m^3}\right)
\end{eqnarray} 
The normalisation constant $A$ is chosen such that 
\begin{eqnarray} 
N = \int n \; d^3\!r = \iint f_{RKD}\; d^3\!v d^3\!r 
\end{eqnarray} 
still holds, so that 
\begin{eqnarray} 
S_{RKD} &=& -k_B N \, \ln\left(\frac{n A h^3}{m^3}\right) \nonumber\\
        & & -k_B N \int {g}_{RKD}\, \ln\left(1 + \frac{v^2}{\kappa\Theta^2}\right)^{-\kappa-1} d^3\!v
            \nonumber\\
        & & +k_B N \frac{\alpha^2}{\Theta^2} \int g_{RKD} v^2\; d^3\!v \nonumber\\
        & & +k_B N
\end{eqnarray} 
The two remaning integrals are (i) independent of location, (ii) finite functions of the 
parameters $\alpha$ and $\kappa >0$, and (iii) independent of particle number $N$. This allows 
to express the entropy for the RKD as:  
\begin{eqnarray} 
S_{RKD} = k_B N \, \left[\ln\left(\frac{1}{n\lambda_{RKD}^3}\right) + I_1(\kappa,\alpha,\Theta) + 1
            + I_2(\kappa,\alpha,\Theta) \right]
\label{rkd-entropy}
\end{eqnarray} 
where we have defined a generalized thermal de Broglie wavelength $\lambda_{RKD} = h/(m A^{1/3})$, and 
the two functions
\begin{eqnarray} 
&& I_1(\kappa,\alpha,\Theta) = (\kappa+1)\, \int {g}_{RKD}\, \ln\left(1 + \frac{v^2}{\kappa\Theta^2}\right)\, d^3\!v\\
&& I_2(\kappa,\alpha,\Theta) = \frac{\alpha^2}{\Theta^2} \int g_{RKD} v^2\; d^3\!v
\end{eqnarray} 
This is the main result: Since all quantities in the square bracket in Eq.(\ref{rkd-entropy}) are independent
of particle number $N$ the entropy $S_{RKD}$ is proportional to $N$ and, thus, an extensive quantity.

Obviously, the Maxwellian case is obtained in the limit $\kappa\to\infty$ with $\alpha=0$. Then
one has
\begin{eqnarray} 
&& \lambda_{RKD} \to \lambda \\
&& I_1(\kappa\to\infty,0,\Theta) \to 3/2 \\
&& I_2(\kappa\to\infty,0,\Theta) = 0
\end{eqnarray} 
so that $S_{RKD}$ correcty reduces to $S_M$. 

Note that, interestingly, this finding may not apply to the standard $\kappa$-distribution,
which is obtained from Eq.(\ref{rkd-dist}) with $\alpha=0$ for $\kappa > 3/2$. This is 
because in the case $\alpha=0$ the number of non-diverging velocity moments is finite and, thus, 
the entropy definition (\ref{entropy}) may not apply \citep{Balescu-1988}. Consequently,  
this `incompleteness' maybe the reason for the non-extensitivity of entropy for the
standard $\kappa$-distribution.   
\section{Isolated, homogeneous plasmas}
\label{isolated}
\begin{figure}
\includegraphics[width = 0.95\columnwidth]{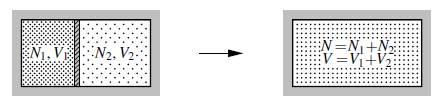}
\caption{Two at first separated plasmas (left box) are eventually occupying the 
         same total volume (right box). Adapted from \cite{Goeke-2010}.}
\label{gibbspara}
\end{figure}

To further elucidate the entropy formula for the RKD let us consider the case of 
two plasma volumes $V_1$ and $V_2$ filled with $N_1$ and $N_2$ particles of the same 
species, see the left box in Figure~\ref{gibbspara}. The two plasmas are in equilibrium with each other, i.e., have 
the same temperature and pressure, and, thus, also the same number density $n = N_1/V_1 = N_2/V_2$.
These plasmas are then mixing and, eventually, fill the total volume $V = V_1 + V_2$ with
$N = N_1 + N_2$ particles, see the right box in Figure~\ref{gibbspara}.
As before, we first briefly recapitulate the case of two Maxwellian plasmas and,
afterwards, that of two RKD plasmas. In both cases we consider thermal equilibrium,
i.e., plasmas of the same temperature. 
%
\subsection{Two Maxwellian plasmas}
For the entropies of the individual plasmas, it is in shown in standard textbooks, e.g., \cite{Goeke-2010} that 
one has with formula (7) and $n = N/V$ the relation 
\begin{eqnarray} 
S_{M,1} + S_{M,2} = S_{M,1+2} =  k_B (N_1 + N_2) \left[\ln\left(\frac{V_1 + V_2}{(N_1 + N_2) \lambda^3}\right)+\frac{5}{2}\right] 
\end{eqnarray} 
This is valid, because with the given constraints, one also has 
\begin{eqnarray} 
\frac{N_2}{V_2} = \frac{N_1}{V_1} = \frac{(N_1/V_1) (V_1 + V_2)}{V_1 + V_2} 
                = \frac{(N_1/V_1) V_1 + (N_2/V_2) V_2)}{V_1 + V_2} = \frac{N_1 + N_2}{V_1 + V_2} 
\label{densities}
\end{eqnarray} 
Under these assumptions, in (local) equilibrium the sum of the entropy of two Maxwellian plasma systems with identical 
particles in separate volumes is equal to the entropy of the mixed plasma filling the total volume. 
%
\subsection{Two RKD plasmas}
First, it is important to note the fact that equal temperature $T_{RKD}$ and equal pressure $p_{RKD}$ for
two plasmas described with RKDs implies that the two distributions have the same $\kappa$- as well as
$\alpha$-values. Second, given that $T_{RKD} = p_{RKD}/(n k_B)$ \cite{Scherer-etal-2017}, they also 
have the same number density, implying that Eq.(\ref{densities}) also holds for two RKD plasmas under the given 
constraints.   

Then, upon introducing the abbreviation 
$F(\kappa,\alpha,\Theta) = I_1(\kappa,\alpha,\Theta) + 1 + I_2(\kappa,\alpha,\Theta)$ in Eq.(\ref{rkd-entropy}), one has 
\begin{eqnarray} 
S_{RKD,1} = k_B N_1 \left[\ln\left(\frac{V_1}{N_1 \lambda_{RKD}^3}\right)+F(\kappa,\alpha,\Theta)\right] \\
S_{RKD,2} = k_B N_2 \left[\ln\left(\frac{V_2}{N_2 \lambda_{RKD}^3}\right)+F(\kappa,\alpha,\Theta)\right] 
\end{eqnarray} 
The same formula yields for the situation when the plasmas have merged: 
\begin{eqnarray} 
S_{RKD,1+2} = k_B (N_1 + N_2) \left[\ln\left(\frac{V_1 + V_2}{(N_1 + N_2) \lambda_{RKD}^3}\right)+F(\kappa,\alpha,\Theta)\right]
\end{eqnarray} 
Exploiting Eq.(\ref{densities}) again results in the finding 
\begin{eqnarray} 
S_{RKD,1} + S_{RKD,2} &=& k_B N_1 \left[\ln\left(\frac{V_1}{N_1 \lambda_{RKD}^3}\right)+F(\kappa,\alpha,\Theta)\right] 
                        + k_B N_2 \left[\ln\left(\frac{V_2}{N_2 \lambda_{RKD}^3}\right)+F(\kappa,\alpha,\Theta)\right] \nonumber\\
                  &=& k_B (N_1+N_2) \left[\ln\left(\frac{V_1+V_2}{(N_1+N_2) \lambda_{RKD}^3}\right)+F(\kappa,\alpha,\Theta)\right]\nonumber\\ 
                  &=& S_{RKD,1+2}
\end{eqnarray} 
Consequently, entirely analogous to the case of two Maxwellian plasmas, one finds that the RKD entropy is an extensive
quantity. Again, as noted in section~\ref{bg-entropy}, this is not necessarily including the case
$\alpha=0$, i.e., the standard $\kappa$-distribution, for which non-extensitivity of entropy has been 
shown \cite[e.g.][]{Tsallis-1988, Leubner-Voeroes-2005}.
%
\section{Spatially inhomogeneous plasmas}
\label{sec-ihs}
We consider an example from space plasma physics, which is not only the origin of $\kappa$-distributions, 
but an area of their frequent application. The subsonic solar wind in the so-called inner heliosheath, i.e., the
region between the shock transition that terminates the supersonic expansion of the solar wind and the heliopause
that separates the solar from the interstellar plasma, is a spatially inhomogeneous plasma. While, due to the 
subsonic flow, incompressibilty is nearly fulfilled \cite{Fahr-Siewert-2015} and, thus, the density is constant,
both the hydrodynamic bulk velocity and the kinetic velocity distributions of suprathermal particles are functions
of the spatial coordinates. The corresponding proton and electron constituents have recently been treated on the 
basis of standard $\kappa$-distributions
\cite{Fahr-etal-2014,Fahr-etal-2016, Fahr-etal-2017}. The evolution of the proton distribution function was described
by deriving a hydrodynamical differential equation for the parameter $\kappa$ as a function of position. 

Depending on the absence or presence of sources or sinks of energy, one can distinguish an isentropic and a non-isentropic 
case, respectively. Starting with the former, we consider the first law of thermodynamics 
\begin{equation}
T dS = dU + p dV
\end{equation}
where the new quantities $U$ and $p$ denote the internal energy and pressure of the plasma component, and 
$V$ is the volume of a plasma parcel. If the flow isentropic and isothermal along a flowline with coordinate 
$s$, i.e.\ $dS/ds = 0$ and $dTd/ds = 0$, the work done by the pressure at the expansion of a moving
plasma volume $\Delta V$ is the only reason to change its internal energy $U$. Consequently, one has
\begin{equation}
\frac{d}{ds}(\varepsilon \cdot \Delta V)  + \frac{d}{ds}\left(p\Delta V\right) = 0
\end{equation}
where we have introduced the energy density $\varepsilon = \Delta U/\Delta V = 3 p/(4\pi)$, and the
second equality follows from its moment definition. This reduces to 
\begin{equation}
\frac{d}{ds}\left(p\Delta V\right) = 0
\end{equation}
Using $p = n k_B T$, Eq.(\ref{kappa-temp}) and incompressibility, i.e.\ $dn/ds = 0$ one obtains
\begin{equation}
\frac{d}{ds}\left[\Theta^{2} \frac{\kappa}{\kappa-3/2} \Delta V\right] = 0 
\end{equation}
and, thus, 
\begin{equation}
{\Theta}^{2}\frac{\kappa}{\kappa-3/2} \Delta V = const
\end{equation}
Particle conservation implies $n u_{sw} \Delta V = const$ along a given streamline (where $u_{sw}$ is
the solar wind plasma convection speed), which for constant $n$ reduces to $u_{sw} \Delta V = const$. so that 
\begin{equation}
u_{sw}^{-1} \Theta^{2}\frac{\kappa}{\kappa-3/2} = const
\end{equation}
For $\Theta=const$ this reproduces the results obtained by \cite{Fahr-etal-2016} for the case of no sources
or sinks and vanishing velocity diffusion: a constant convection speed $u_{sw}$ along a streamline implies constant
$\kappa$ and an increasing (decreasing) speed results in an increasing (decreasing) $\kappa$. While constant $\kappa$
yields, via Eq.(\ref{kappa-temp}), both constant temperature $T$ and constant reference speed $\Theta$,  
increasing (decreasing) $\kappa$ would translate into decreasing (increasing) temperature \cite{Scherer-etal-2018}.
This, however, is excluded here by the above assumption of $T=const$, so that $\Theta$ cannot be considered constant.
The latter combination is equivalent to Kappa~A as discussed in section~\ref{sec-kappa-dist}. Given that the
heliosheath is not isothermal \citep{Richardson-Wang-2012}, however, it is more likely that Kappa~B is the
appropriate choice.
 
In case the flow along the streamlines developes non-isentropically due to presence of energy
sources and sinks, i.e.\ if the entropy of the fluid changes with the flow line element $s$, 
one has to consider the following relation for the entropy per volume $\hat{S} = nS$
\begin{equation}
\frac{d\hat{S}}{ds} = \frac{1}{T}\frac{d\hat{Q}}{ds} = \frac{1}{U T}\frac{d\hat{Q}}{dt} 
\end{equation}
which follows from $dS = dQ/T$ with $\hat{Q} = nQ$ and the incompressibility condition $n=const$.  
The newly introduced quantitiy $dQ$ describes changes of the internal energy of a comoving
volume element $dV$. As discussed in \cite{Fahr-etal-2016}, these changes are due to (i) velocity
diffusion with a diffusion coefficient proportional to $v^2$ and (ii) the so-called magnetic cooling.  
The related changes are, as calculated in \cite{Fahr-etal-2016} for
standard $\kappa$-distributions, proportional to the thermal pressure:
\begin{equation}
\frac{dQ}{dt} = 10 D_0 p(s) - \frac{4 U(s)}{3 B(s)}\frac{dB}{ds}p(s)
\end{equation}
with $D_0$ denoting a diffusion constant and $B$ the strength of the magnetic field.
Since the same holds for the temperature via $T = p/(n k_B)$, the change of entropy density along a streamline 
\begin{equation}
\frac{d\hat{S}}{ds} = \frac{n k_B}{U p(s)}\frac{dQ}{dt} 
                    = n k_B \left(\frac{10 D_0}{U} - \frac{4}{3 B(s)}\frac{dB}{ds}\right) 
\end{equation}
is independent of $\kappa$. Consequently, one obtains
\begin{equation}
\hat{S}(s) = \hat{S}(s_{0}) + n k_B\int\limits_{s_{0}}^{s}\left(\frac{10 D_0}{U}
                                                              - \frac{4}{3 B(s)}\frac{dB}{ds}\right)\, ds
           = \hat{S}(s_{0}) + n k_B\left[10 D_0 \int\limits_{s_{0}}^{s}\frac{1}{U}\, ds 
                                               - \frac{4}{3}\,\ln\left(\frac{B(s)}{B_0}\right)\right]
\end{equation}
which describes the change of entropy density along a given streamline. Note, first, that this expression is
via $\hat{S}(s_0)$ still depending on $\kappa$ and, second, that for other velocity diffusion models and
other distribution functions (e.g., the RKD) also the entropy density change will depend on $\kappa$.   
%
\section{Summary and conclusions}
\label{conclusions}
Starting from the general (Boltzmann-)Gibbs definition, we derived, first, a formula for the entropy 
of a spatially homogeneous plasma whose constituents can be modelled on the basis of the regularized
$\kappa$-distribution. Second, we have demonstrated that for these distribution functions entropy 
is, analogous to a Maxwellian plasma, an extensive quantity. And, third, we have discussed the change
of entropy (density) along streamlines in an incompressible, but otherwise inhomogeneous flow.

In conclusion, we state that within the framework of regularized $\kappa$-distributions entropy can be
defined in such a way that it maintains -- in difference to the case of the standard $\kappa$-distributions --
its additivity, which appears mandatory in view of the fundamental laws of thermodynamics.
%
\begin{acknowledgements}
We are grateful for support from the Deutsche Forschungsgemeinschaft (DFG) via the grants SCHE~334/9-2,
SCHL~201/35-1, and from FWO-Vlaanderen (Grant GOA2316N). We also appreciate the support from the
International Space Science Institute (ISSI) for hosting the international ISSI team on {\it Kappa Distributions: 
From Observational Evidences via Controversial Predictions to a Consistent Theory of Suprathermal
Space Plasmas}, which triggered many fruitful discussions that were
beneficial for the work presented here. We are also grateful for helpful comments by Christian R\"oken.
\end{acknowledgements}
%
%
\end{document}